\documentclass[
   ,draft            
  ]
  {aipproc}

\layoutstyle{8x11single}

\def\ltsima{$\; \buildrel < \over \sim \;$}
\def\la{\lower.5ex\hbox{\ltsima}}            
\def\gtsima{$\; \buildrel > \over \sim \;$}
\def\ga{\lower.5ex\hbox{\gtsima}}            

\begin{document}

\title 
      [Gamma-ray emitting AGN and GLAST]
      {Gamma-ray emitting AGN and GLAST}


\author{P. Padovani}{
  address={European Southern Observatory, Karl-Schwarzschild-Str. 2, 
  85748 Garching bei M\"unchen, Germany},
}

\begin{abstract}
I describe the different classes of Active Galactic Nuclei (AGN) and the
basic tenets of unified schemes. I then review the properties of the
extragalactic sources detected in the GeV and TeV bands, showing that the
vast majority of them belong to the very rare blazar class. I further
discuss the kind of AGN GLAST is likely to detect, making some predictions
going from the obvious to the likely, all the way to the less probable.
\end{abstract}

\maketitle


\section{The Active Galactic Nuclei Zoo}

Active Galactic Nuclei (AGN) are extragalactic sources, in some cases
clearly associated with nuclei of galaxies (although generally the host
galaxy light is swamped by the nucleus), whose emission is dominated by
non-stellar processes in some waveband(s).

Based on a variety of observations, we believe that the inner parts of AGN
are not spherically symmetric and therefore that emission processes are
highly anisotropic \cite{a93,up95}. The current AGN paradigm includes a
central engine, almost certainly a massive black hole, surrounded by an
accretion disk and by fast-moving clouds, which under the influence of the
strong gravitational field emit Doppler-broadened lines. More distant
clouds emit narrower lines. Absorbing material in some flattened
configuration (usually idealized as a torus) obscures the central parts, so
that for transverse lines of sight only the narrow-line emitting clouds are
seen and the source is classified as a so-called "Type 2" AGN. The
near-infrared to soft-X-ray nuclear continuum and broad-lines, including
the UV bump typical of classical quasars, are visible only when viewed
face-on, in which case the object is classified as a "Type 1" AGN. In
radio-loud objects, which constitute $\approx 10\%$ of all AGN, we have the
additional presence of a relativistic jet, likely perpendicular to the disk
(see Fig. 1 of \cite{up95}).

This axisymmetric model of AGN implies widely different observational
properties (and therefore classifications) at different aspect
angles. Hence the need for "Unified Schemes" which look at intrinsic,
isotropic properties, to unify fundamentally identical (but apparently
different) classes of AGN. Seyfert 2 galaxies are though to be the "parent"
population of, and have been "unified" with, Seyfert 1 galaxies, whilst
low-luminosity (Fanaroff-Riley type I [FR I]\cite{fr74}) and
high-luminosity (Fanaroff-Riley type II [FR II] ) radio galaxies have been
unified with BL Lacs and radio quasars respectively \cite{up95}. In other
words, BL Lacs are thought to be FR I radio galaxies with their jets at
relatively small ($\la15 - 20^{\circ}$) angles w.r.t. the line of sight.
Similarly, we believe flat-spectrum radio quasars (FSRQ) to be FR II radio
galaxies oriented at small ($\la 15^{\circ}$) angles, while steep-spectrum
radio quasars (SSRQ) should be at angles in between those of FSRQ and FR
II's ($15^{\circ} \la \theta \la 40^{\circ}$; a spectral index value
$\alpha_{\rm r} = 0.5$ at a few GHz [where $f_{\nu} \propto \nu^{-\alpha}$]
is usually taken as the dividing line between FSRQ and SSRQ). BL Lacs and
FSRQ, that is radio-loud AGN with their jets practically oriented towards
the observer, make up the blazar class. Blazars, as I show below, play a
very important role in $\gamma$-ray astronomy and it is therefore worth
expanding on their properties.

\subsection{Blazars}\label{blazars}

\begin{figure}
  \includegraphics[height=0.5\textheight]{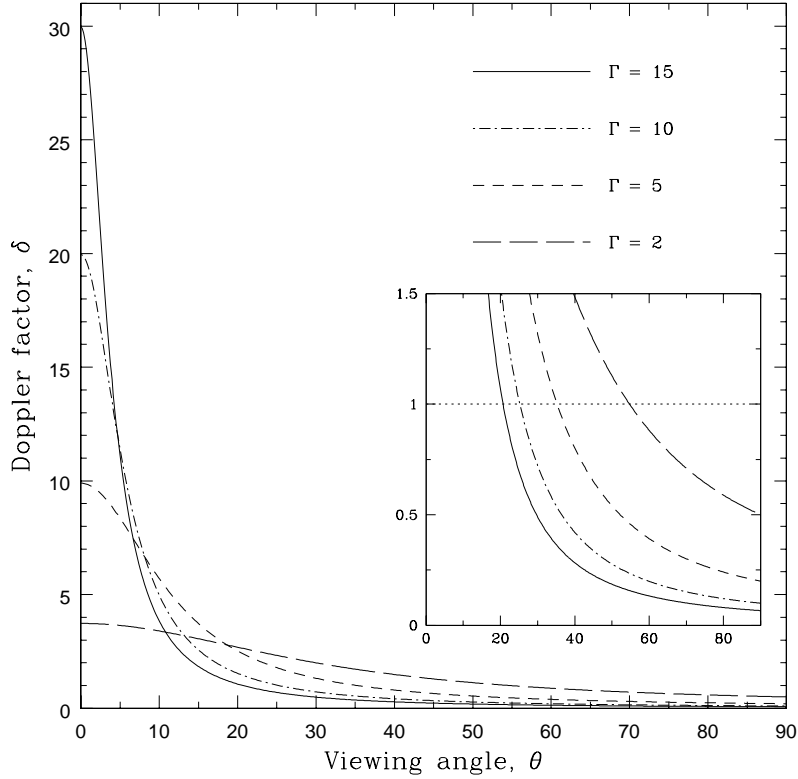}
  \caption{The dependence of the Doppler factor on viewing angle. Different
  curves correspond to different Lorentz factors $\Gamma$. The expanded
  scale on the inset shows the angles for which
  $\delta=1$. }\label{pad:fig1}
\end{figure}

Blazars are the most extreme variety of AGN. Their signal properties
include irregular, rapid variability, high polarization, core-dominant
radio morphology (and therefore flat [$\alpha_{\rm r} \la 0.5$] radio
spectra), apparent superluminal motion, and a smooth, broad, non-thermal
continuum extending from the radio up to the $\gamma$-rays \cite{up95}.
Blazar properties are consistent with relativistic beaming, that is bulk
relativistic motion of the emitting plasma at small angles to the line of
sight, which gives rise to strong amplification and collimation in the
observer's frame. Adopting the usual definition of the relativistic Doppler
factor $\delta = [\Gamma (1 - \beta \cos \theta)]^{-1}$, $\Gamma =
(1-\beta^2)^{-1/2}$ being the Lorentz factor, $\beta = v/c$ being the ratio
between jet speed and the speed of light, and $\theta$ the angle w.r.t. the
line of sight, and applying simple relativistic transformations, it turns
out that the observed luminosity at a given frequency is related to the
emitted luminosity in the rest frame of the source via $L_{\rm obs} =
\delta^p L_{\rm em}$ with $p \sim 2 - 3$. For $\theta \sim 0^{\circ}$,
$\delta \sim 2 \Gamma$ (Fig.~\ref{pad:fig1}) and the observed luminosity
can be amplified by factors 400 -- 10,000 (for $\Gamma \sim 10$ and $p \sim
2 - 3$, which are typical values). That is, for jets pointing almost
towards us the emitted luminosity can be overestimated by up to four orders of
magnitude. For more typical angles $\theta \sim 1/\Gamma$, $\delta \sim
\Gamma$ and the amplification is $\sim 100 - 1,000$.

In a nut-shell, blazars can be defined as sites of very high energy
phenomena, with bulk Lorentz factors up to $\Gamma \approx 30$ \cite{co07}
(corresponding to velocities $\sim 0.9994$c) and photon energies reaching
the TeV range (see below).

Given their peculiar orientation, blazars are very rare. Assuming that the
maximum angle w.r.t. the line of sight an AGN jet can have for a source to
be called a blazar is $\sim 15^{\circ}$, only $\sim 3\%$ of all radio-loud
AGN, and therefore $\approx 0.3\%$ of all AGN, are blazars. For a $\sim 1 -
10\%$ fraction of galaxies hosting an AGN, this implies that only 1 out of
$\approx 3,000 - 30,000$ galaxies is a blazar!

Blazar spectral energy distributions (SEDs) are usually explained in terms
of synchrotron and inverse Compton emission, the former dominating at lower
energies, the latter being relevant at higher energies. Blazars have a
large range in synchrotron peak frequency, $\nu_{\rm peak}$, which is the
frequency at which the synchrotron energy output is maximum (i.e., the
frequency of the peak in a $\nu - \nu f_{\nu}$ plot). Although the
$\nu_{\rm peak}$ distribution appears now to be continuous, it is still
useful to divide blazars into low-energy peaked (LBL), with $\nu_{\rm
peak}$ in the IR/optical bands, and high-energy peaked (HBL) sources, with
$\nu_{\rm peak}$ in the UV/X-ray bands \cite{pg95}. The location of the
synchrotron peaks suggests in fact a different origin for the X-ray
emission of the two classes. Namely, an extension of the synchrotron
emission responsible for the lower energy continuum in HBL, which display
steep ($\alpha_{\rm x} \sim 1.5$) X-ray spectra \cite{wol98}, and inverse
Compton (IC) emission in LBL, which have harder ($\alpha_{\rm x} \sim 1$)
spectra \cite{pad04}. This distinction applies almost only to BL Lacs, as
most known FSRQ are of the low-energy peak type and, therefore, with the
X-ray band dominated by inverse Compton emission.  Very few ``HFSRQ'' (as
these sources have been labelled), i.e., FSRQ with high (UV/X-ray energies)
$\nu_{\rm peak}$ are in fact known. Moreover, $\nu_{\rm peak}$ for all
these sources (apart from one) appears to be $\sim 10 - 100$ times smaller
than the values reached by BL Lacs (see \cite{pad07} for a review).

\section{The GeV and TeV skies}

Before moving on to GLAST we need to assess the present status of the
$\gamma$-ray sky. I do this first at GeV and then TeV energies.

The third EGRET catalogue \cite{har99} includes 271 sources ($E > 100$
MeV), out of which 95 were identified as extragalactic (including 28 lower
confidence sources). Further work \cite{mat01,so03,so04}, which provided
more identifications, allows us to say that EGRET has detected at least
$\sim 130$ extragalactic sources (since a large fraction of sources is
still unidentified), all of them AGN apart from the Large Magellanic
Cloud. Furthermore, all the AGN are radio-loud and $\sim 97\%$ of them are
blazars, with the remaining sources including a handful of radio galaxies
(e.g., Centaurus A, NGC 6251).  Most of the blazars are FSRQ, in a ratio
$\sim 3/1$ with BL Lacs. Finally, $\sim 80\%$ of the BL Lacs are LBL and
the few HBL are all local ($z < 0.12$). As all of the FSRQ are also of the
LBL type, $\sim 93\%$ of EGRET detected blazars are of the low-energy peak
type.

The situation at TeV energies is at first order similar to that in the GeV
band, with some significant differences. All confirmed extragalactic TeV
sources are radio-loud AGN and include 16 BL Lacs and one radio galaxy
(M87) (a starburst galaxy is also a possible TeV source)
\cite{maz07,al07b}. That is, the blazar fraction is $\sim 94\%$. Unlike the
GeV band, however, no FSRQ is detected and all but one BL Lacs are
HBL. This is due to the fact that in HBL the very high-energy flux is
higher than in LBL, as both peaks of the two humps in their SED are shifted
to higher frequencies.

The fact that the GeV and TeV skies are dominated by blazars seems to be at
odds with these sources being extremely rare (see previous section). The
explanation has to be found in the peculiar properties of the blazar class
and rests on the fact that blazars are characterized by:

\begin{enumerate}
\item high-energy particles, which can produce GeV and TeV photons; 
\item relativistic beaming, to avoid photon-photon collision and amplify the
flux;
\item strong non-thermal (jet) component. 
\end{enumerate}

Point 1 is obvious. We know that in some blazars synchrotron emission
reaches at least the X-ray range, which reveals the presence of high-energy
electrons which can produce $\gamma$-rays via inverse Compton emission
(although other processes can also be important: e.g., \cite{ce07}). Point
2 is vital, as otherwise in sources as compact as blazars all GeV photons,
for example, would be absorbed through photon-photon collisions with target
photons in the X-ray band (see, e.g., \cite{ma92}). Beaming means that the
intrinsic radiation density is much smaller than the observed one and
therefore $\gamma$-ray photons manage to escape from the source. The flux
amplification in the observer's frame makes also the sources more easily
detectable. Point 3 is also very important. $\gamma$-ray emission is
clearly non-thermal (although we still do not know for sure which processes
are responsible for it) and therefore related to the jet component. The
stronger the jet component, the stronger the $\gamma$-ray flux.

\section{GLAST and AGN}

We can know ask which (and how many) AGN GLAST will detect. This I describe
in the following, in decreasing order of "obviousness".

\subsection{Blazars}

Given that blazars are well know $\gamma$-ray sources, GLAST will certainly
detect many flat-spectrum radio quasars and BL Lacs. How many exactly
depends on a variety of factors. These include blazar evolution and
intrinsic number density (which can to some extent be estimated from deep
surveys in other bands), their duty cycle in the $\gamma$-ray band (as we
know that EGRET was detecting mostly sources in outburst), and their SED
(see below). Finally, any prediction will have not to violate the
extragalactic $\gamma$-ray background.

To get an order of magnitude estimate, I make the following simple
assumptions: a) EGRET has detected 130 blazars, which is likely to be a
lower limit given the still unidentified sources; b) the number counts are
Euclidean, that is $N(>S) \propto S^{-1.5}$, where $S$ is the flux density;
this is a very likely upper limit as we know that, after the initial steep
rise, number counts of extragalactic sources tend to flatten out at lower
fluxes; c) GLAST is 30 times more sensitive than EGRET. The total number of
blazars GLAST will detect over the whole sky is then $\la 20,000$. This
corresponds to $\la 0.5$ objects/deg$^{2}$, which, interestingly enough, is
the surface density of blazars down to $\sim 50$ mJy at 5 GHz in the Deep
X-ray Radio Blazar Survey (DXRBS) \cite{pad07b}. Note also that by means of
Monte Carlo simulations a value around 5,000 has been predicted all-sky
(extrapolating from the high Galactic latitude value of \cite{gio07}; see
also \cite{gio07b}).

\subsection{HBL}

As discussed above, EGRET has detected very few blazars of the high-energy
peak type (HBL). This is because the EGRET band was sampling the "valley"
between the two (synchrotron and IC) humps in their SED. A look at the SED
of some of the TeV detected HBL \cite{al06,al07,tav01} shows that many, if
not all, of them should be easily detected by GLAST.

\subsection{Radiogalaxies}

Unified schemes predict that the "parent" population of blazars is made up
of radio-galaxies, a much more numerous class (by a factor $\approx 30$ for
a dividing angle between the two classes $\sim 15^{\circ}$). However, at
large angles w.r.t. the line of sight, jet emission is not only
not-amplified but actually de-amplified. Fig. \ref{pad:fig1} shows that for
typical Lorentz factors $\delta < 1$ for viewing angles $\ga 20 -
30^{\circ}$. This implies that radio-galaxies on average are weaker sources
(by factors $\approx 1,000$) than blazars, in all bands. And indeed, the
handful of GeV/TeV-detected radio-galaxies are all local ($z < 0.02$).

Large scale, that is kpc-scale jet emission, as opposed to the small,
pc-scale, one, is also unlikely to be relevant in the $\gamma$-ray band for
the bulk of radio-galaxies \cite{sta03,sa07}.

However, the radio-galaxy cause might not be totally lost. It has been
proposed that blazar jets are structured or decelerated. The first scenario
\cite{ghi05}, which ties in with Very Long Baseline Interferometry (VLBI)
observations of limb brightening \cite{gir04}, suggests the presence of a
fast spine surrounded by a slower external layer. In the other case
\cite{geo05}, which tries to reconcile the low $\delta$ values from VLBI
observations of TeV BL Lacs with the high values inferred from SED modeling
of the same sources, the jet is supposed to decelerate from a Lorentz
factor $\Gamma \sim 20$ down to $\Gamma \sim 5$ over a length of $\sim 0.1$
pc. In both instances the presence of the two velocity fields implies that
each of the two components sees an enhanced radiation field produced by the
other. The net result is that IC emission gets boosted and therefore the
GeV flux is higher than that predicted in the simpler case of an
homogeneous jet (at the price of having a larger number of free
parameters).  Assuming that the $\gamma$-ray/radio flux ratio observed for
the three GeV/TeV-detected radio-galaxies sources is typical, at least 10
3CR radio-galaxies should to be detected by GLAST \cite{ghi05}.

Note that some Broad Line Radio Galaxies (BLRG), which are Type 1 sources
in which the jet is at angles intermediate between those of blazars and
radio-galaxies, are also likely to be detected by GLAST
\cite{gra07a,gra07b}.

\subsection{Radio-Quiet AGN}

The large majority of AGN are of the radio-quiet type, that is they are
characterized by very weak radio emission, on average $\sim 1,000$ times
fainter than in radio-loud sources. Radio-quiet does not mean radio-silent,
however, and the nature of radio emission in these sources is still
debated. Two extreme options ascribe it either to processes related to
star-formation (synchrotron emission from relativistic plasma ejected from
supernovae) or to a scaled down version of the non-thermal processes
associated with energy generation and collimation present in radio-loud
AGN.  In the latter case, one would expect also radio-quiet AGN to be (faint)
$\gamma$-ray sources.  Assuming their GeV flux to scale roughly as the
radio flux this would be, on average, a factor $\approx 30$ below the GLAST
detection limit. Detection might be possible, however, for the (few) high
core radio flux radio-quiet AGN. Even a negative detection, supported by
detailed calculations, could prove very valuable in constraining the nature
of radio-emission in these sources.

\section{Summary}\label{pad:sum}

The main conclusions are as follows: 

\begin{enumerate}
\item{Blazars, even though they make up a small minority of AGN, dominate
the $\gamma$-ray sky;}
\item{GLAST will certainly detect "many thousand" blazars, with the exact
number being somewhat model dependent;}
\item{GLAST will most likely detect "many" high-energy peaked blazars,
which have so far escaped detection at GeV energies due to the fact that
EGRET was sampling the "valley" between the two (synchrotron and IC) humps
in their spectral energy distribution;}
\item{GLAST will possibly detect a "fair" number of radio-galaxies;}
\item{GLAST might also detect some radio-quiet AGN, depending on the nature
of their radio emission.}
\end{enumerate}

In any case, GLAST will constrain (radio-loud) AGN physics and populations,
as described very well at this conference!

\begin{theacknowledgments}
It is a pleasure to thank Paolo Giommi for useful discussions and Annalisa
Celotti for reading the manuscript.
 \end{theacknowledgments}

\end{document}